\documentclass[%
reprint,
notitlepage,
 showpacs,preprintnumbers,
 aps,
pre,
 longbibliography
]{revtex4-1}

\usepackage{amsmath,amssymb,amsfonts}
\usepackage{comment}
\usepackage{graphicx}
\usepackage{textcomp}
\usepackage{xcolor}
\usepackage{color}
\usepackage{dsfont}
\usepackage[percent]{overpic}
\usepackage{tikz, pgfplots}

\newcommand{\iS}{{\bar{i}}}
\newcommand{\jS}{{\bar{j}}}

\newcommand{\nS}{{\bar{n}}}
\newcommand{\Dt}{{\Delta t}}
\newcommand{\Dg}{{\Delta}}
\newcommand{\ie}{i.~e.}

\newcommand{\imdea}{\affiliation{Instituto Madrile\~{n}o de Estudios Avanzados en Nanociencia (IMDEA-Nanociencia), Cantoblanco, 28049 Madrid, Spain}}%
\newcommand{\uam}{\affiliation{Departamento de Qu\'imica, Universidad Aut\'onoma de Madrid, M\'odulo 13, Madrid, 28049, Spain}}

\newcommand{\emugr}{\affiliation{Department of Electromagnetism, University of Granada, Granada, 18071, Spain}}%

\newcommand{\nudt}{\affiliation{National University of Defense Technology. 137 Yanwachi, Changsha, Hunan, 410073. People's Republic of China.}}%

\begin{document}

\title{Numerical simulation of knotted solutions for Maxwell equations}
%\thanks{The work described in this paper and the research leading to these results have been supported by the Spanish MINECO, EU FEDER under Project TEC2016-79214-C3-3-R (MINECO, Spain), and the Alhambra-LFT contract with AIRBUS (Spain).}

\author{Antonio M. Valverde}\emugr
\author{Luis D. Angulo}\emugr
\author{Juan J. Omiste}\imdea\uam
\author{M. R. Cabello}\emugr
\author{Salvador G. Garc\'ia}\emugr
\author{Jianshu Luo}\nudt

\date{\today}

\begin{abstract}
In this work, we use the finite differences in time domain (FDTD) numerical method to compute and assess the validity of Hopf solutions, or hopfions, for the electromagnetic field equations. 
In these solutions, field lines form closed loops characterized by different knot topologies which are preserved during their time evolution.
Hopfions have been studied extensively in the past from an analytical perspective but never, to the best of our knowledge, from a numerical approach.
The implementation and validation of this technique eases the study of more complex cases of this phenomena; e.g. how these fields could interact with materials (e.g. anisotropic or non-linear), their coupling with other physical systems (e.g. plasmas), and also opens the path on their artificial generation by different means (e.g. antenna arrays or lasers).
\end{abstract}

\maketitle

\section{Introduction}
\label{sec:introduction}

Hopfions are a family of localized solutions for the  electromagnetic field Maxwell equations in which field lines are closed, forming knotted topologies which are preserved when evolved in time~\cite{Irvine2008,Lechtenfeld2018,Cameron2018}. 
Beyond their intrinsic mathematical interest, these solutions may also contribute to several branches of Physics.
Some authors have proposed that they play a key role in the phenomena known as \textit{ball lightning}~\cite{Ranada1996,Ranada2000} or as exotic quantum mechanical solutions that describe the electron at a fundamental level, predicting some of its properties~\cite{Lee2018, Ranada1998a,Ranada_1992}. 

In the past, hopfions have been studied exclusively from an analytical perspective \cite{Irvine2008, Irvine2010, Kedia2013, Ranada1989, Kawaguchi2008}. However, to the best of our knowledge, they have never been simulated numerically. 
These simulations can be of interest for several reasons. First, because hopfions are demonstrated to exist, not only from purely analytical arguments, but from the direct numerical resolution of the elemental Maxwell's curl equations. Additionally, a validated numerical approach opens many possibilities to study more complicated variants of this phenomena, e.g. the study of their interactions with anisotropic or non-linear  materials, with other hopfions~\cite{Arrayas2017a}, their coupling with other physical equations, or the possibility of generating them by means of antenna arrays or lasers; a technique which was proposed as a way for their physical realization but has not been accomplished yet \cite{Irvine2008}.

The accuracy assessment of hopfion numerical solutions is a necessary step to address certain physical problems which are not feasible analytically. To this end, in this work we use the finite-difference time-domain (FDTD) method \cite{Yee1966, Taflove2005}, a proven and robust method which is ubiquitous in computational electrodynamics, and which is possibly the optimal approach given the spatial and time scales involved. The input of the method is an initial known analytical hopfion solution and the obtained numerical evolution results are then compared with the expected analytical solution. Different metrics are proposed as tools to asses the validity of this approach.

This work is organized as follows: first, in Sec.~\ref{sec:background} we give a theoretical background, focusing on the construction of hopfions and their helicity conservation property. Second, we briefly describe the FDTD method. Next, in Sec.~\ref{sec:results_and_validation} we describe the propagation of the hopfion based on the numerical simulations as well as the conservation of the helicity as a benchmark. In this context, we also introduce a metric to quantify the error in the propagation. Finally, in Sec.~\ref{sec:conclusions_and_outlook} we summarize the main conclusion of this study along with possible extensions and applications.

\section{Background}
\label{sec:background}

\subsection{Theory}
\label{sec:theory}

Hopfions were proposed in 1989 by  Ra\~{n}ada~\cite{Ranada1989}. In that work, he formulated a particular solution for Maxwell's equation in which all field lines are closed and form a torus which deforms over time while, at the same time, preserves its topology. This result was then extended and categorized as part of a family of solutions characterized by an arbitrary number of mathematical \textit{knots},~\ie, embeddings of a circle in a 3-dimensional Euclidean space~\cite{Irvine2008,Kawaguchi2008}. In this regard, Ra~\~{n}ada's torus has a circumferential core which within this family of solutions corresponds to the knot known as {\em the unknot}. However, with the exception of Ra\~{n}ada's hopfion, these generalized hopfion topologies did not preserve over time. It was not until 2013 when Kedia et al.~[\onlinecite{Kedia2013}] brought to light an analytical construction which allows to formulate a whole family of topology preserving knotted solutions for Maxwell's equations.  

It is widely accepted that light knots must be null fields as a necessary condition to preserve knot topology~\cite{Kedia2013, Besieris2009}. The properties of the null fields are
\begin{enumerate}
	\item Electric and magnetic fields are perpendicular at every point: $\boldsymbol{E}\bot\boldsymbol{B}$
	\item They have the same magnitude: $|\boldsymbol{E}| = |\boldsymbol{B}|$.
\end{enumerate}
Assuming this hypothesis, we may apply the {\em Bateman's method} to build null fields, and hence, hopfion solutions. First, we define the Riemann-Silberstein vector $\boldsymbol{F}$,
\begin{equation}\label{eq:Riemann-Silberstein}
    \boldsymbol{F} = \boldsymbol{E}+i\boldsymbol{B}
\end{equation}
where $\boldsymbol{B}$ and $\boldsymbol{E}$ represent the magnetic and electric field, respectively. Note that we use the electromagnetic natural units in which the vacuum electric permittivity, magnetic permeability, and speed of light ($\varepsilon_0$, $\mu_0$ and $c$, respectively) are equal to one.

Bateman's method proves that every Riemann-Silberstein vector corresponding to a null field can be written as,
\begin{equation}\label{eq:Bateman}
    \boldsymbol{F} = \nabla\alpha\times\nabla\beta
\end{equation}
where $\alpha$ and $\beta$ can be any complex function as long as they meet the following condition:
\begin{align}
    \nabla\alpha\times\nabla\beta=i\left(\frac{\partial\alpha}{\partial t} \nabla\beta - \frac{\partial\beta}{\partial t}\nabla\alpha\right)
\end{align}
Kedia et al. \cite{Kedia2013} found these specific expressions for $\alpha$ and $\beta$:
\begin{equation}\label{eq:alpha}
    \alpha=\left(\frac{r^2-t^2-1+2iz}{r^2-(t-i)^2}\right)^p
\end{equation}
\begin{equation}\label{eq:beta}
    \beta=\left(\frac{2(x-iy)}{r^2-(t-i)^2}\right)^q
\end{equation}
where $p$ and $q$ must be positive coprime integers which lead us to different kinds of knots.
Note that the expressions \eqref{eq:alpha} and \eqref{eq:beta} have been obtained assuming an arbitrary distance ($l_0$) and time ($t_0=l_0/c$) units which set the hopfion scale.

The knotness of a hopfion can be characterized by its {\it magnetic} and {\it electrical helicity}, $h_B$ and $h_E$ respectively \cite{Irvine2008,Arrayas2017}, defined as
\begin{equation}
    \label{eq:hb_he}
    h_B(\boldsymbol{B})=\int_D \boldsymbol{A}\cdot\boldsymbol{B} \, \mathrm{d}^3\boldsymbol{r}, \,\,\, h_E(\boldsymbol{E})=\int_D \boldsymbol{C}\cdot\boldsymbol{E} \, \mathrm{d}^3\boldsymbol{r},
\end{equation}
where $\boldsymbol{B}=\nabla\times\boldsymbol{A}$, $\boldsymbol{E}=\nabla\times\boldsymbol{C}$ and $D$ represents the domain.
Note that as the helicity is a measurable quantity, it must be gauge independent. 

Let us prove that the magnetic helicity is invariant under gauge transformations, i.e., the magnetic helicity for $\boldsymbol{A}+\nabla f$, noted as $\bar{h}_B(\boldsymbol{B}$), is equal to $h_B(\boldsymbol{B})$. We compute $\bar{h}_B(\boldsymbol{B})$ using Eq.~\eqref{eq:hb_he}
\begin{eqnarray}
    \nonumber
    \bar{h}_B(\boldsymbol{B})&=&\int_D (\boldsymbol{A}+\nabla f)\cdot\boldsymbol{B}\,\mathrm{d}^3\boldsymbol{r}={h}_B(\boldsymbol{B})+\int_D \nabla f\cdot\boldsymbol{B}\,\mathrm{d}^3\boldsymbol{r}\\
    \nonumber
    &=&h_B(\boldsymbol{B})+\int_D \nabla\cdot(f \boldsymbol{B})\,\mathrm{d}^3\boldsymbol{r}-\int_D f\nabla\cdot \boldsymbol{B}\,\mathrm{d}^3\boldsymbol{r}
    \\
    \label{eq:hb_gauge}
    &=&h_B(\boldsymbol{B})+\int_{\partial D} f\cdot \boldsymbol{B}\,\mathrm{d}\boldsymbol{S}=h_B(\boldsymbol{B}),
\end{eqnarray}
where we have used that $\nabla \cdot \boldsymbol{B}=0$ and imposed that the magnetic field vanishes in the boundary of $D$, i.e. $\boldsymbol{B}|_{\partial D}=0$. Using a similar procedure we obtain that the helicity of the electric field is also preserved.

Now, we prove that $h_B$ and $h_E$ do not change in time for null fields. The time derivative of the magnetic helicity
\begin{eqnarray}
    \nonumber
    \partial_t h_B(\boldsymbol{B})&=& \int_D\partial_t\boldsymbol{A}\boldsymbol{B}\,\mathrm{d}^3\boldsymbol{r}+\int_D\boldsymbol{A}\partial_t\boldsymbol{B}\,\mathrm{d}^3\boldsymbol{r}=\\
    \label{eq:hb_partial_t}
    &=&-\int_D \boldsymbol{E}\cdot\boldsymbol{B}\,\mathrm{d}^3\boldsymbol{r}+\int_D\boldsymbol{A}\nabla\times\partial_t\boldsymbol{A}\,\mathrm{d}^3\boldsymbol{r}
\end{eqnarray}
then, we use that~\cite{Griffiths1999},
\begin{equation}
    \nabla(\boldsymbol{U}\times\boldsymbol{V})=(\nabla\times \boldsymbol{U})\cdot \boldsymbol{V}-(\nabla\times\boldsymbol{V})\cdot \boldsymbol{U},
\end{equation}
and we set $\boldsymbol{U}=\boldsymbol{A}$ and $\boldsymbol{V}=\partial_t\boldsymbol{A}$. Thus, Eq.~\eqref{eq:hb_partial_t} reads as
\begin{eqnarray}
    \nonumber
    \partial_t &h_B&(\boldsymbol{B})=-\int_D \boldsymbol{E}\cdot\boldsymbol{B}\,\mathrm{d}^3\boldsymbol{r}-\int_D\nabla(\boldsymbol{A}\times\partial_t\boldsymbol{A})\,\mathrm{d}^3\boldsymbol{r}\\
    \nonumber
    &=&\int_D\partial_t\boldsymbol{A}\nabla\times\boldsymbol{A}\,\mathrm{d}^3\boldsymbol{r}=\\
    \label{eq:hb_partial_t_zero}
    &=& -2\int_D \boldsymbol{E}\cdot\boldsymbol{B}\,\mathrm{d}^3\boldsymbol{r}-\int_{\partial D}\boldsymbol{A}\times\partial_t\boldsymbol{A}\,\mathrm{d}\boldsymbol{S}=0,
\end{eqnarray}
where we have used that the null fields fulfilled that $\boldsymbol{E}\cdot\boldsymbol{B}=0$ and $\partial_t\boldsymbol{A}=-\boldsymbol{E}$ vanishes in the boundary $\partial D$. Using a similar procedure we can prove that the electric helicity is constant over time.

\subsection{The Finite-Differences in Time Domain method}
\label{sec:fdtd}
The Yee FDTD scheme~\cite{Yee1966} numerically solves Maxwell curl equations by replacing the space and time derivatives by finite differences. Any unknown field component can be advanced a time step using the ones at adjacent space positions. To obtain an optimized algorithm, fields are arranged strategically on the center of the edges and faces of a cubic cell of size $\Dg$; a configuration known as Yee's cell \cite{Yee1966}. For instance, to evolve the $E_z$ component a time step $\Dt$ we obtain the following formula in free-space,
\begin{align}\label{eq:FDTD:YEE:1D:EH}
    &\left.E_{z}\right|_{\iS,\jS,k}^{\nS} =
    \left.E_{z}\right|_{\iS,\jS,k}^{\nS-1}
    \nonumber \\
    & + \frac{\Dt}{\Dg} \left( {\left.H_{y}\right|_{i,\jS,k}^{n}-
    \left.H_{y}\right|_{i+1,\jS,k}^{n}} -{\left.H_{x}\right|_{\iS,j,k}^{n}+
    \left.H_{x}\right|_{\iS,j+1,k}^{n}} \right)
\end{align}
where $i,j,k$ are integer numbers identifying each cell and time step at a time step $n$; barred indices indicate the addition of a half-step, e.g. $\iS = i + 1/2$, and $B=\mu H$.

In order to compute a simulation, Maxwell's equations have to be propagated in a finite box endowed with boundary conditions. As our aim is to simulate a hopfion which is supposed to be isolated in an infinite space, we must set reflection-less boundary conditions which absorb all the energy exiting the domain. The most widely used method in FDTD for this purpose is the perfectly matched layer (PML), which is reported to perform with less than -80 dB of attenuation for reflections, independently of the angle of incidence of impinging waves~\cite{Gedney1996}. 

\section{Results and validation}
\label{sec:results_and_validation}

In this section, we study the time propagation of the numerical hopfion solutions, cross-validated with analytical results. The time-domain nature of the FDTD method allows us to visualize different snapshots of the time evolution. The initial time step is set to $t=-1.5t_0$, the computational domain is a cubic box of $17.5l_0\times 17.5l_0\times 17.5l_0$, being the spatial step  $\Dg=0.025\,l_0$ and a temporal step of $\Dt=0.8\Dg/\sqrt{3}$.  All the results shown in this work were obtained using an Intel(R) Core(TM) i7-4710MQ personal computer with 16 GB of RAM.
We focus our discussion on two types of hopfions obtained by setting $\{p=1, q=1\}$ and $\{p=2, q=3\}$ in expressions~\eqref{eq:alpha} and \eqref{eq:beta}. 

\begin{figure}[b] %ELECTRIC FIELDS FOR {11}
	\centering
	\includegraphics[width=\linewidth]{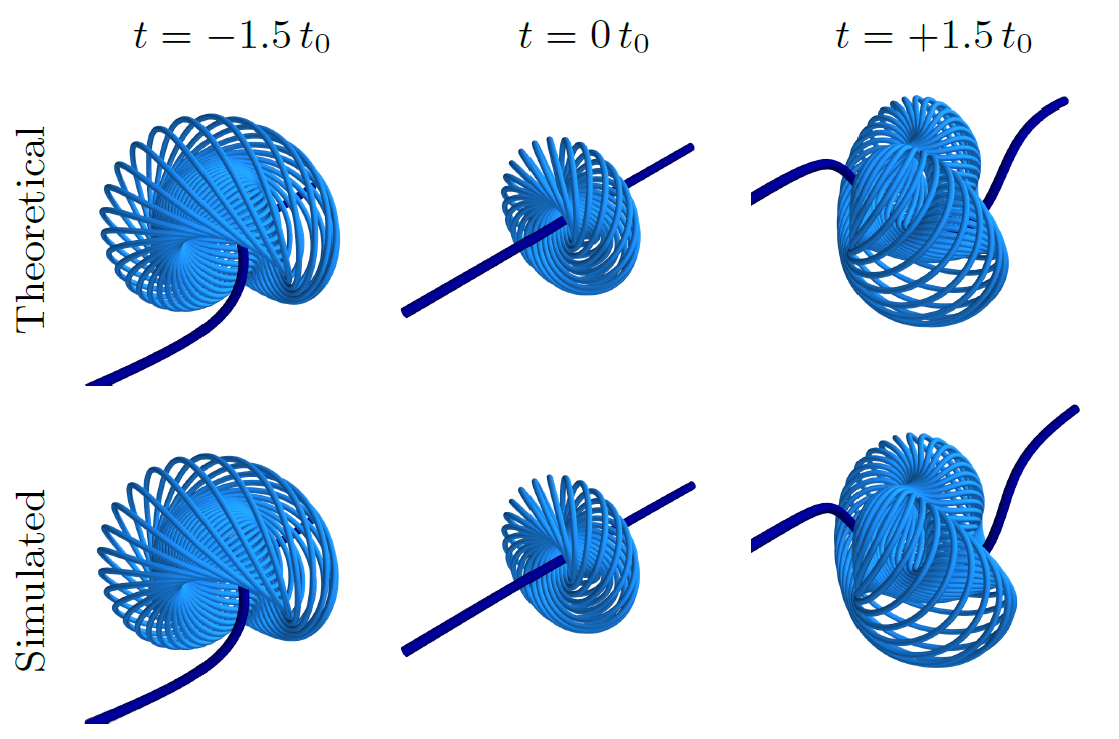}
	\caption{\label{fig:fig1} Analytical (first row) and computed (second row) electric field lines for hopfion \{1,1\}. Dark lines correspond to the only field line that does not close on itself.}
\end{figure}

\subsection{Hopfion's dynamics}

\begin{figure}[b] %MAGNETIC FIELDS FOR {11}
	\centering
	\includegraphics[width=\linewidth]{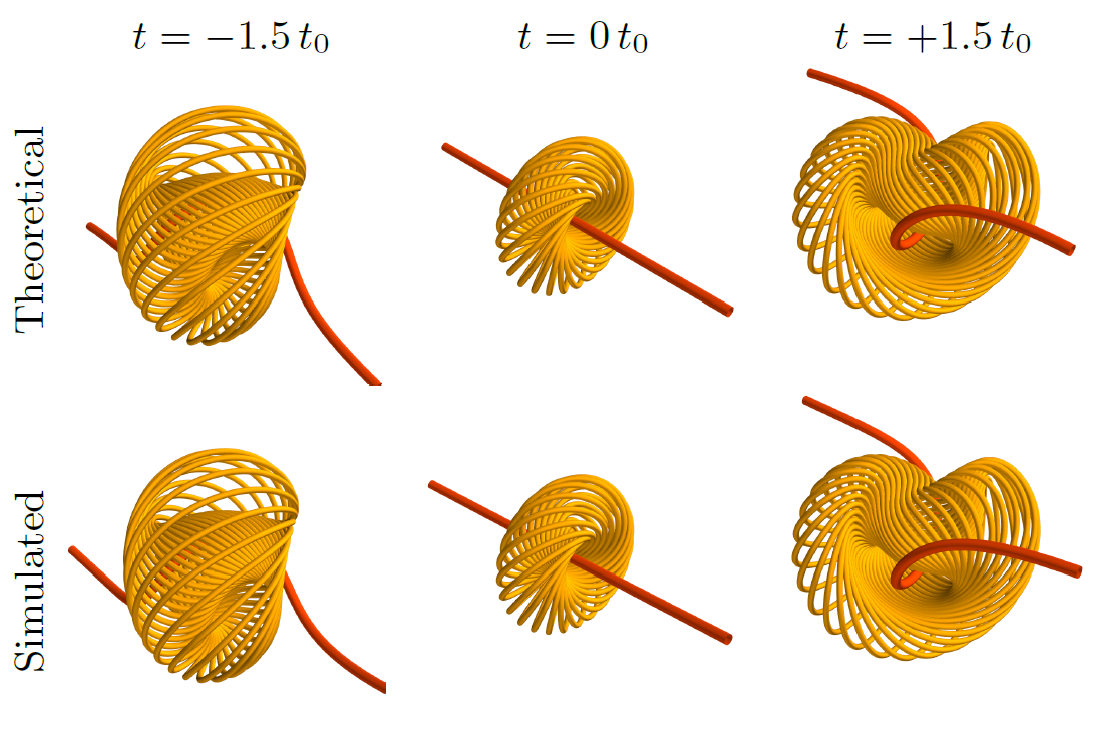}
	\caption{\label{fig:fig2}  For hopfion \{1,1\} analytical (first row) and computed (second row) magnetic field lines. Dark lines correspond to the only field line that does not close on itself.}
\end{figure}

First, we analyze the hopfion \{1,1\}, also known as Ra\~{n}ada's torus or the unknot. In Fig.~\ref{fig:fig1} and ~\ref{fig:fig2} we show the analytical and numerically computed electric and magnetic field lines, respectively, at three different times. Let us remark that the torus shape-like of field lines are similar for the electric and magnetic field. Furthermore, the computational results reproduce very accurately the analytical ones, being indistinguishable in the scale of the figure. 
\begin{figure}[b] %POYNTING FIELDS FOR {11}
	\centering
	\includegraphics[width=\linewidth]{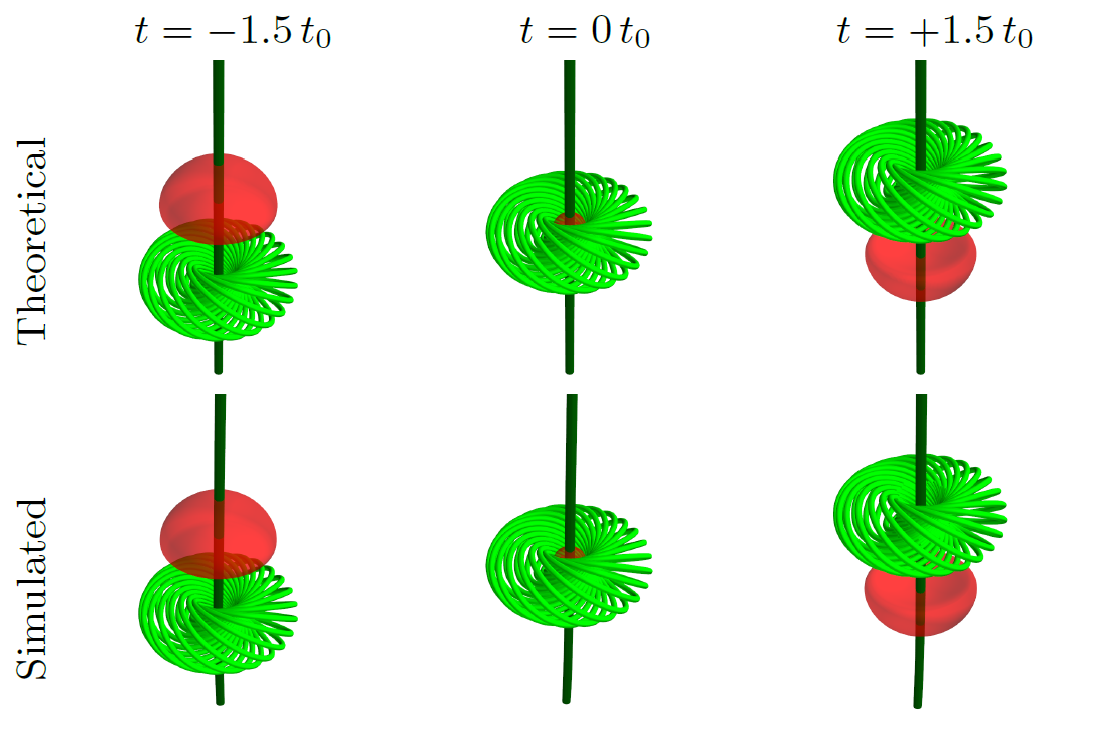}
	\caption{\label{fig:fig3}  For hopfion \{1,1\}, analytical (first row) and computed (second row) Poynting field lines. Dark lines correspond to the only field line that does not close on itself. Red surface correspond to an area in which Poynting vector has a constant magnitude (half of its maximum value).}
\end{figure}
On the other hand, in Fig.~\ref{fig:fig4} and~\ref{fig:fig5} we show the analytical and computed hopfion \{2,3\} solution for the electric and magnetic fields as a function of time. As in the previous case, it topology structure,~\ie, the trefoil knot, is preserved during the time propagation. However, we see that the field lines of simulated hopfion at $t=1.5t_0$ are closer than in the theoretical simulation,~\ie, this simulation tends to stick the field lines. This error occurs because of the complexity of hopfion \{2,3\}, whose structure is more tangled than hopfion \{1,1\}. However, this error can be suppressed by using a more dense computational grid. Even though, this error is not remarkable, as we will discuss in more detail in Sec.~\ref{sec:error_propagation}.

In figures~\ref{fig:fig3} and~\ref{fig:fig6} show Poynting vector field lines for hopfion \{1,1\} and \{2,3\}, respectively. Interestingly, in both cases we can appreciate that they correspond to the {\em unknot}, even for hopfion \{2,3\}, whose electric and magnetic field correspond to the {\em trefoil knot}. Moreover, for both hopfions we may note that the torus defined by the Poynting vector field moves from bottom to top of the $Z$ axis without deformation while the electromagnetic energy moves from top to bottom. The main difference we appreciate is the energy distribution. If we now compare the simulated and theoretical results we may see that they are virtually identical, even considering that interpolations of the simulation results were necessary to obtain these field lines.

\begin{figure}[h] %ELECTRIC FIELDS FOR {23}
	\centering
	\includegraphics[width=\linewidth]{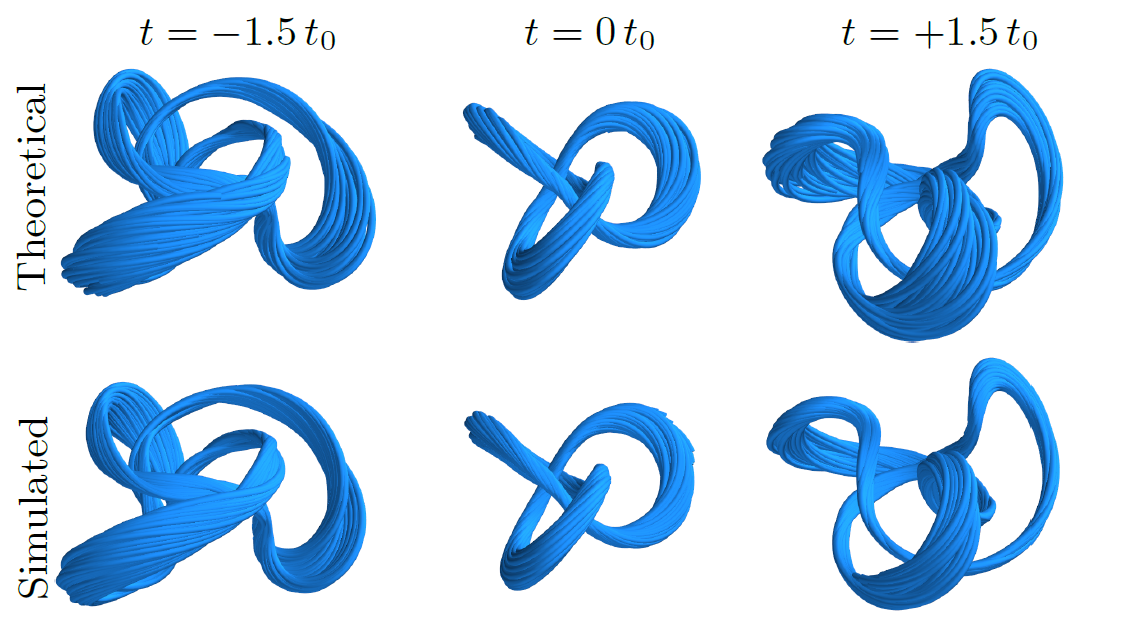}
	\caption{\label{fig:fig4} For hopfion \{2,3\}, analytical (first row) and computed (second row) electric field lines.}
\end{figure}

\begin{figure}[h] %MAGNETIC FIELDS FOR {23}
	\centering
	\includegraphics[width=\linewidth]{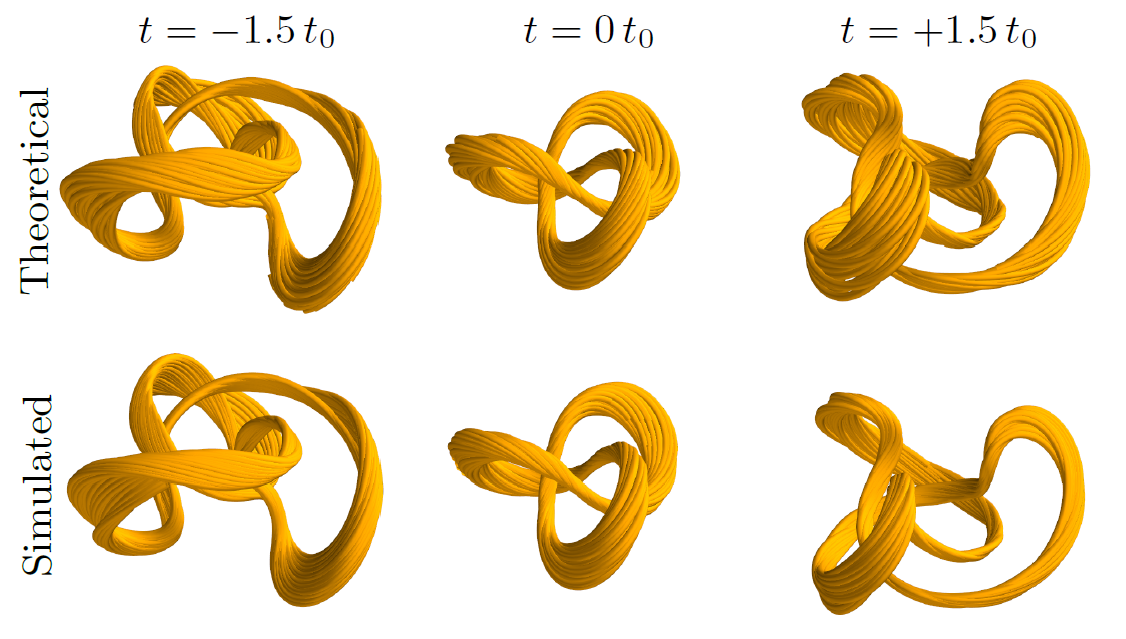}
	\caption{\label{fig:fig5} For hopfion \{2,3\}, analytical (first row) and computed (second row) magnetic field lines.}
\end{figure}

\begin{figure}[h] %POYNTING FIELDS FOR {23}
	\centering
	\includegraphics[width=\linewidth]{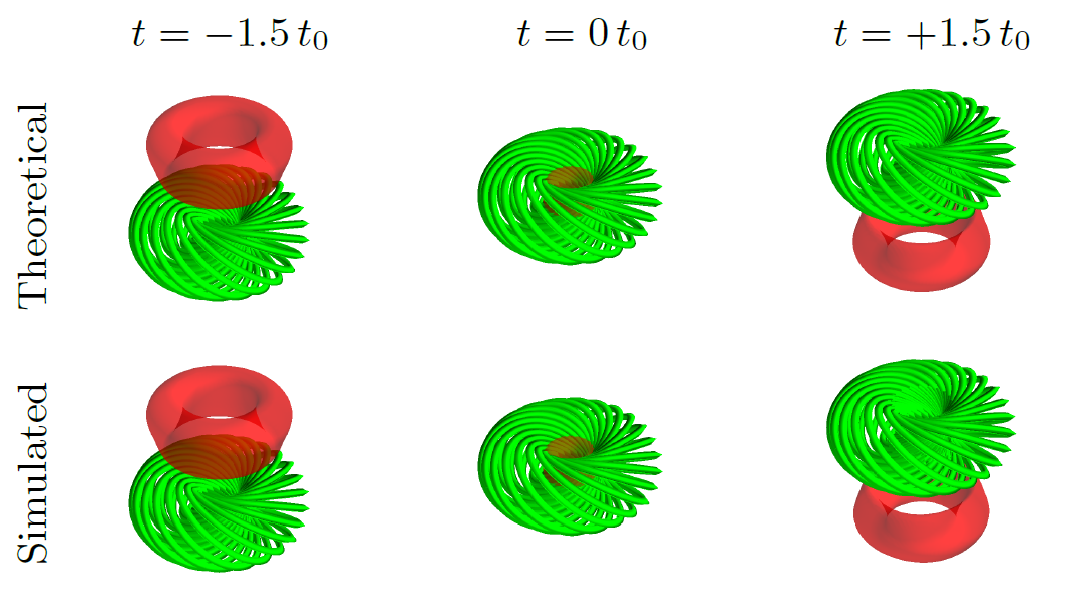}
	\caption{\label{fig:fig6} For hopfion \{2,3\}, analytical (first row) and computed (second row) Poynting field lines. Red surface correspond to an area in which Poynting vector has a constant magnitude (half of its maximum value).}
\end{figure}

At first glance, the knotness is conserved during the numerical propagation, as in the theoretical solution. In order to quantify the conservation of knotness, we now investigate the helicity, which is related to the topology of the solution, since it only takes a non-zero value if the topology of the magnetic (electric) field lines are not trivial~\cite{Ranada_1992}. In particular we analyze the time propagation of the magnetic helicity, $h_B$, without loss of generality, since the results are equivalent for $h_E$. To do so, we define the normalized derivative of $h_B$ as
\begin{align}\label{eq:normalized_dt_hb}
    \left(\partial_t h_B\right)_\text{norm.} = \dfrac{
        \partial_t h_B
    }{
        \int_D\left|\bf{E}\right|\cdot\left|\bf{B}\right|
        \,\mathrm{d}^3 \bf{r}
    }.
\end{align}
As we have proven in Eq.~\eqref{eq:hb_partial_t_zero}, the helicity associated to the magnetic and electric field is a conserved quantity in a hopfion, thus, we expect that $\left(\partial_t h_B\right)_\text{norm.}=0$ during the propagation. 

\begin{figure}[t]
    \centering
    \includegraphics[width=\linewidth]{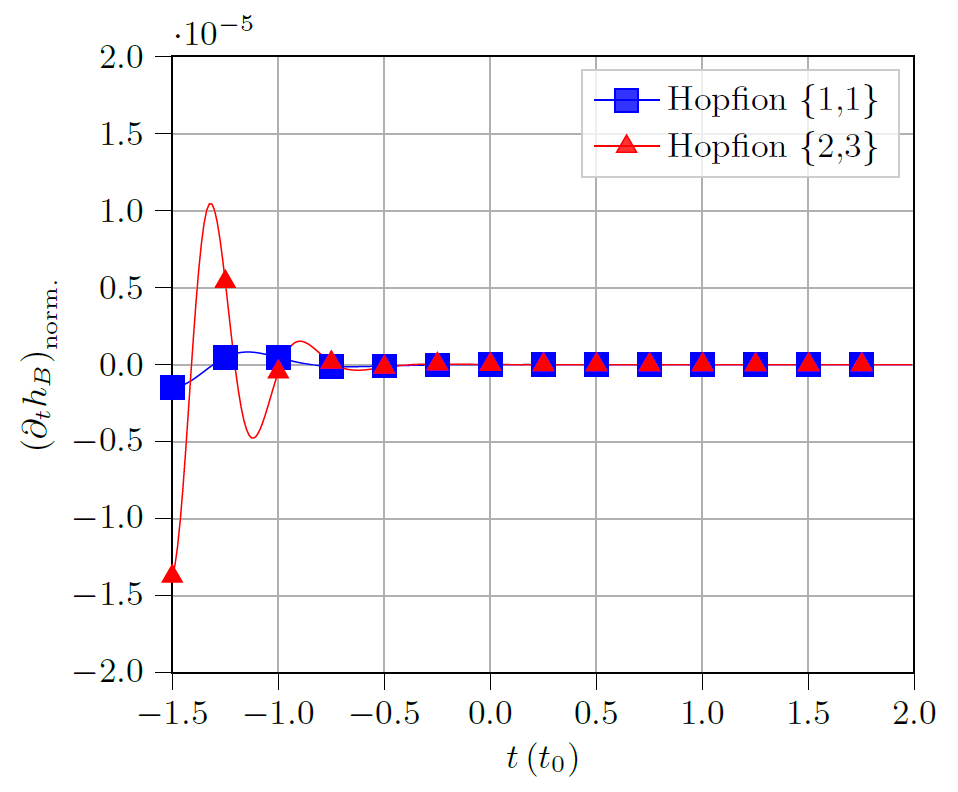}
    \caption{Representation of normalized $\partial_t h_B$ over time for hopfion \{1,1\} and \{2,3\}.}
    \label{fig:fig7}
\end{figure}

In order to validate our method, we plot $\left(\partial_t h_B\right)_\text{norm.}$ as a function of time in Fig.~\ref{fig:fig7}. Note that we compute $\left(\partial_t h_B\right)_\text{norm.}$ for an inset with a side of $10l_0$ centered in the computational domain. For both cases we find that $\left(\partial_t h_B\right)_\text{norm.}$ at $t=-1.5t_0$ is slightly different from zero and continues oscillating before converging to zero exponentially. This behaviour at initial times is associated to modes unsupported by FDTD, caused by the discretization of the analytical hopfion used as initial condition.
However, these modes do not play a role in long term dynamics as FDTD makes $\left(\partial_t h_B\right)_\text{norm.}$ converge to zero after a few oscillations. This process takes around $2t_0$, which is a short interval compared to the temporal scale associated with the size of the simulation box. 

\subsection{Error propagation}
\label{sec:error_propagation}

In this section, we use the following metric to assess the numerical error of our methodology 
\begin{equation}
\text{Err}(U)\big|_{i,j,k}^{n} = \left| \frac{ U_{theor}\big|_{i,j,k}^{n} -U_{sim}\big|_{i,j,k}^{n} } {U_{theor}\big|_{i,j,k}^{n}} \right|,
\label{eq:error}
\end{equation}
 where $U$ corresponds to a field, $i,j,k$ to a point in the grid and $n$ to the $n$th time step.

\begin{figure*}
    \centering
    \includegraphics[width=\linewidth]{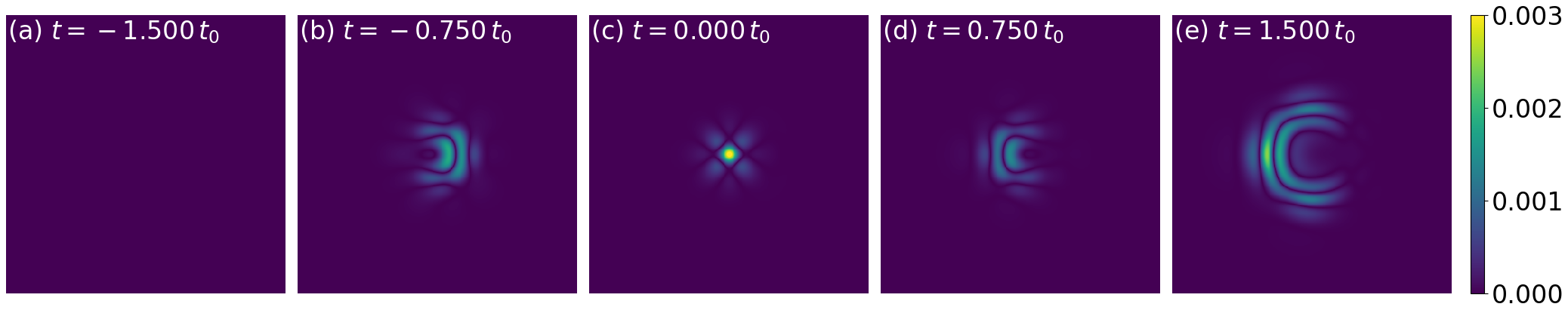}
    \caption{For the hopfion \{1, 1\}, simulation error of the $E_y$ component at the cross-section $Y=0$.}
    \label{fig:fig8}
\end{figure*}

\begin{figure*}
    \centering
    \includegraphics[width=\linewidth]{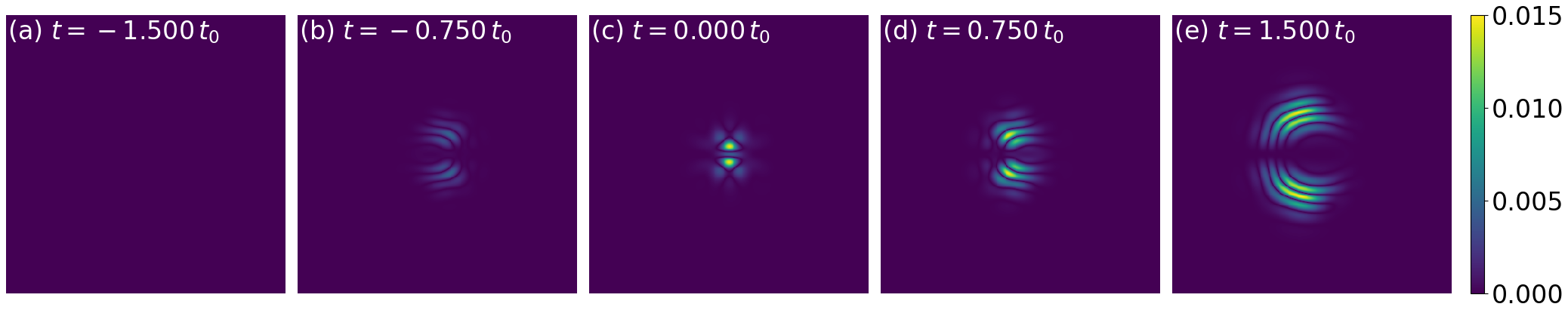}
    \caption{For the hopfion \{2, 3\}, simulation error of the $E_y$ component at the cross-section $Y=0$.}
    \label{fig:fig9}
\end{figure*}

In Figs.~\ref{fig:fig8} and \ref{fig:fig9}, we show $\text{Err}(U)\big|_{i,j,k}^{n}$ on the plane $Y=0$ for the $y$-component of $\boldsymbol{E}$ for the hopfions $\{1,1\}$ and $\{2,3\}$, respectively. Note that we represent the error only for one vector component in order to avoid any numerical artifact which could be caused by the interpolation necessary due to the staggered nature of the FDTD algorithm. This metric evaluates the relative error with respect to the theoretical value of each component at every point, which allows us to obtain the percentage error. This is expected to be higher where the theoretical value is zero. In these points, the most tiny differences can cause the relative error to go up to infinity. For visualization purposes, we have chosen to show the numerical error for a component in which the theoretical value is never zero, this being $E_y$ at plane $Y=0$.
 The figures correspond to an inset of $10l_0$ from a computational domain of $17.5l_0$.

First of all, we realize that the error for hopfion \{1,1\} is smaller than for hopfion \{2,3\} which can be attributed to the numerical dispersion due to the higher spatial and temporal variations of the latter solution. We observe that the error propagates in every direction, being negligible at $t=-1.5t_0$ since it corresponds to the first iteration. After that, the error increases, spreading around the center of the grid, as we see for instance on Fig.~\ref{fig:fig8}(b). At $t=1.5t_0$ the error is less that $2\%$ for hopfion \{2,3\} and even lower for hopfion \{1,1\}, with a maximum value of $0.3\%$.

The numerical dispersion error has no impact on the helicity, a basic property of hopfions, which demonstrates the suitability of the FDTD to propagate this type of solutions. Specifically, the FDTD preserves the component $\mathbf{E}\cdot \mathbf{B}$ by construction, being an ideal method to simulate null fields, as it is the case.
 
\section{Conclusions and outlook}
\label{sec:conclusions_and_outlook}

We have demonstrated that FDTD is a viable alternative to simulate different kinds of light knots. Specifically, taking hopfion \{1,1\} as a benchmark, we have shown that  the FDTD is an efficient and accurate method to propagate this solution of Maxwell's equation. Using the same grid we have reproduced accurately the dynamics of a more tangled hopfion, in particular hopfion \{2,3\}, which shows that FDTD is suitable to simulate this kind of structures. Besides, we have shown that the helicity variation converges to zero in a short time, which is a proof that the topology is preserved during the propagation. Furthermore, it is important to note that FDTD method would allow for improved results, if necessary, by using a finer discretization.

This work opens new lines of investigation, since it paves the way to investigate the hopfions beyond the analytical expressions~\eqref{eq:alpha} and~\eqref{eq:beta}, which are too complex to solve analytically in many physical systems. For example, the FDTD will allow to investigate their generation and confinement as well as the interaction of hopfions among them or with other structures such as metal, meta-materials or plasma, among many others. Moreover, the numerical simulations can be used to design experimental setups to produce hopfions in the laboratory. This kind of experiments are of great interest, since they may measure the ball lightning, which has been hypothesized to be hopfions or hopfions linked to plasma.

\begin{acknowledgments}
%The authors are within the Dept. of Electromagnetism, Universidad de Granada, Fuentenueva s/n, 18071 Granada, Spain. 
J.J.O acknowledges the funding from the project FULMATEN-CM (Ref: Y2018/NMT-5028) funded by the Programme of R\&D Activities of Comunidad de Madrid (Spain) and the Social European Fund.
\end{acknowledgments}

\end{document}